`

# Spin Relaxation in Single Layer and Bilayer Graphene

Wei Han and R. K. Kawakami[†]

Department of Physics and Astronomy, University of California, Riverside, CA 92521

[†] e-mail: roland.kawakami@ucr.edu

**Abstract:**

We investigate spin relaxation in graphene spin valves and observe strongly contrasting behavior for single layer graphene (SLG) and bilayer graphene (BLG). In SLG, the spin lifetime ($\tau_s$) varies linearly with the momentum scattering time ($\tau_p$) as carrier concentration is varied, indicating the dominance of Elliot-Yafet (EY) spin relaxation at low temperatures. In BLG, $\tau_s$ and $\tau_p$ exhibit an inverse dependence, which indicates the dominance of Dyakonov-Perel spin relaxation at low temperatures. The different behavior is due to enhanced screening and/or reduced surface sensitivity of BLG, which greatly reduces the impurity-induced EY spin relaxation.

PACS numbers: 85.75.-d, 81.05.ue, 72.25.Rb, 72.25.Hg

`

Graphene is an attractive material for spintronics due to the possibility of long spin lifetimes arising from low intrinsic spin-orbit coupling and weak hyperfine coupling [1-5]. However, Hanle spin precession experiments in graphene spin valves report spin lifetimes that are orders of magnitude shorter than expected theoretically [6-12]. This has prompted theoretical studies of the extrinsic sources of spin relaxation such as impurity scattering [13], ripples [5], and substrate effects [14]. Experimentally, several studies have investigated spin relaxation including the roles of impurity scattering [7, 11] and graphene thickness [15]. Recently, it has been shown that ferromagnet (FM) contact-induced spin relaxation is responsible for the short spin lifetimes observed in experiments [12]. Therefore, high quality tunneling contacts are necessary to suppress the contact-induced effects for systematic investigations of spin relaxation in graphene.

In this Letter, we perform systematic studies of spin relaxation in single layer graphene (SLG) and bilayer graphene (BLG) spin valves with tunneling contacts. The dependence of spin lifetime on temperature and carrier concentration (tuned by gate voltage) reveals rather different spin relaxation mechanisms in the two systems. In SLG, the temperature dependence shows similar trends of the spin lifetime and momentum scattering time, and the low temperature gate voltage dependence shows a strong linear scaling of the two quantities. This indicates the dominance of Elliot-Yafet (EY) spin relaxation, which most likely comes from impurity scattering. In BLG, the temperature dependence and low temperature gate voltage dependence show a nearly inverse relationship between the spin lifetime and momentum scattering time. This indicates the dominance of Dyakonov-Perel (DP) spin relaxation, which can be generated by ripples in the graphene. The contrasting behaviors of SLG and BLG can be understood as a reduction of the impurity scattering in BLG due to the enhanced screening of the impurity potential and reduced surface sensitivity. This leads to longer spin lifetimes (~6.2 ns, the highest value observed in graphene spin valves to date) and the greater role of DP spin relaxation observed in BLG.

The graphene flakes are mechanically exfoliated from HOPG onto an SiO$_2$ (300 nm thickness)/Si substrate [16]. Gate voltage is applied to the Si substrate to tune the carrier concentration in graphene. SLG and BLG are identified by optical microscopy and Raman spectroscopy [17]. Standard e-beam lithography with PMMA/MMA bilayer resist is used to define the Au and Co electrodes. First, two Au electrodes are put down on the two ends of the graphene. Then a second step of e-beam lithography is used for the Co electrodes, where subsequent angle evaporations of TiO$_2$, MgO, and Co produce the ferromagnetic electrodes with tunneling contacts [12, 18]. In ultrahigh vacuum, 0.12 nm of Ti is deposited at an angle of 9° and is converted to TiO$_2$ by exposure to oxygen partial pressure of $5 \times 10^{-8}$ torr for 30 min. A 0.8 nm layer of MgO is deposited at 9° for the tunnel barrier and 3 nm of MgO is deposited at 0° for a masking layer. 80 nm of Co is deposited at 7° for the ferromagnetic electrodes, which are capped with 5 nm of Al$_2$O$_3$ prior to lift-off. The Ti, MgO, and Al$_2$O$_3$ are deposited by electron beam evaporation and the Co is deposited from a thermal effusion cell. Typically several Co electrodes are fabricated in between the two Au electrodes, but only two Co electrodes are wired up for the nonlocal measurement. The widths of the Co electrodes vary between 80 nm and 300 nm to have different coercivities.

Studies of spin transport and spin relaxation are performed on graphene spin valves consisting of two spin-sensitive Co electrodes (E2, E3) and two Au electrodes (E1, E4). Nonlocal voltages (V$_{NL}$) are measured using lock-in detection with an ac injection current of $I = 1$ μA rms at 13 Hz [18]. For the nonlocal measurement of spin transport, spin-polarized carriers are injected into the graphene at electrode E2 [6, 19-24]. The spins subsequently diffuse to electrode E3 where they are detected as a voltage $V_{NL}$ measured across electrodes E3 and E4. The nonlocal resistance ($R_{NL} = V_{NL}/I$) is measured as a function of in-plane magnetic field (Fig. 1a inset) to detect the spin injection and transport. Figures 1a show the nonlocal magnetoresistance (MR) curves for a typical SLG device (Device A), in which the sharp changes in $R_{NL}$ are due to the magnetization switching of the Co electrodes. The nonlocal MR ($\Delta R_{NL}$) is indicated by the red arrow, which is the magnitude of the sharp change in $R_{NL}$.

The spin lifetime ($\tau_s$), diffusion coefficient ($D$), and spin diffusion length ($\lambda = \sqrt{D\tau_s}$) are determined by the Hanle spin precession measurement (figure 1b inset) [20]. Applying an out-of-plane magnetic field ($H_\perp$) causes the spins to precess as they diffuse from E2 to E3, which results in characteristic Hanle curves as shown for devices A at 300 K (figure 1b). The red circles (black circles) are for the parallel (antiparallel) alignment of the Co magnetizations. $\tau_s$ and $D$ are determined by fitting the Hanle curves with

$$R_{NL} \propto \pm \int_0^\infty \frac{1}{\sqrt{4\pi Dt}} \exp[-\frac{L^2}{4Dt}]\cos(\omega_L t)\exp(-t/\tau_s)dt \qquad (1)$$

where the + (-) sign is for the parallel (antiparallel) magnetization state, $L$ is the spacing between the Co electrodes, $\omega_L = g\mu_B H_\perp/\hbar$ is the Larmor frequency, $g$ is the g-factor, $\mu_B$ is the Bohr magneton, and $\hbar$ is the reduced Planck's constant. For device A, $D = 0.013$ m$^2$/s, $\tau_s = 447$ ps, and $\lambda_G = 2.4$ μm. The $\tau_s$ and $D$ obtained from the Hanle curves are plotted as a function of gate voltage in Figure 1c and 1d for 300K and 4K respectively. At 300 K, there is no obvious correlation between $\tau_s$ and $D$. Interestingly, when the device is cooled to $T = 4$ K, $\tau_s$ and $D$ exhibit a strong correlation, with both quantities increasing with carrier concentration. The correlation of $\tau_s$ and $D$ implies a linear relation between $\tau_s$ and the momentum scattering time, $\tau_p$ ($D \sim \tau_p$ as discussed in refs. [7, 15, 25]). This indicates that at low temperatures the spin scattering is dominated by momentum scattering through the EY mechanism (i.e. finite probability of a spin-flip during a momentum scattering event) [26-28]. This behavior has been observed in five SLG devices.

The temperature dependences of $\tau_s$ and $D$ at different carrier concentrations are shown in Figures 2a and 2b. As the temperature decreases from 300 K to 4 K, $\tau_s$ shows a modest increase at higher carrier densities (e.g. from ~0.5 ns to ~1 ns for $V_g - V_{CNP} = +60$ V) and little variation for lower carrier densities. The temperature dependence of $D$ shows a similar behavior as $\tau_s$. To analyze the relationship between the spin scattering and momentum scattering, we plot $\tau_s$ vs. $D$ for temperatures T = 4 K (figure 2c), T = 10 K (figure 2d), T = 50 K and 100 K (figure 2e), T = 200 K and 300 K (figure 2f), respectively. The main trend is that for lower temperatures, $\tau_s$ scales linearly with $D$, which indicates that an EY spin relaxation mechanism is dominant at lower temperatures (≤100 K). For higher temperatures, $\tau_s$ and $D$ do not follow the linear relationship as shown at low temperatures, which suggests that multiple sources of spin scattering are present. If there is more than one source of EY spin scattering (e.g. impurities of different species, phonons, etc.), the linear relationship between $\tau_s$ and $\tau_p$ does not necessarily hold; for example, with two EY scattering mechanisms obeying $\tau_{s,1}^{-1} = k_1 \tau_{p,1}^{-1}$ and $\tau_{s,2}^{-1} = k_2 \tau_{p,2}^{-1}$, the overall spin relaxation rate $\tau_s^{-1} = \tau_{s,1}^{-1} + \tau_{s,2}^{-1} = k_1 \left( \tau_{p,1}^{-1} + \frac{k_2}{k_1} \tau_{p,2}^{-1} \right)$ is not proportional to the overall momentum scattering rate $\tau_p^{-1} = \tau_{p,1}^{-1} + \tau_{p,2}^{-1}$ except in some special cases (e.g. $k_1 = k_2$, $\tau_{p,1} \ll \tau_{p,1}$, etc.).

Next, we investigate spin relaxation in BLG spin valves, which differs from SLG not just in thickness but also in band structure (linear for SLG, parabolic for BLG) and intrinsic spin-orbit coupling [29, 30]. Figure 3a shows the Hanle curve of the longest observed spin lifetime of 6.2 ns, which is obtained on device B for the charge neutrality point at 20 K. Figure 3a inset shows the Hanle curve at 300 K with best fit parameters of $\tau_s$ = 268 ps. The gate voltage dependences of $\tau_s$ and $D$ at 300 K and 20 K are shown in Figure 3b and 3c, respectively. At 300 K, $\tau_s$ varies from 250 ps to 450 ps as a function of gate voltage and exhibits no obvious correlation with $D$. At 20 K, $\tau_s$ varies from 2.5 ns to 6.2 ns, showing a peak at the charge neutrality point. On the other hand, the gate voltage dependence of $D$ exhibits lower values near the charge neutrality point and increasing values at higher carrier densities. The opposite behaviors of $\tau_s$ and $D$ suggest the importance of DP spin relaxation (i.e. spin relaxation via precession in internal spin-orbit fields) where $\tau_s$ scales inversely with $\tau_p$ [28, 31]. This behavior has been observed in four BLG devices. Figures 3d and 3e show the temperature dependences of $\tau_s$ and $D$, respectively. At low temperatures, $\tau_s$ is enhanced while $D$ is reduced, which is different from SLG where both D and $\tau_s$ increase as temperature decreases for most gate voltages. The opposite trends of the temperature dependences of $\tau_s$ and $D$ suggest the strong contributions of spin relaxation mechanisms of the DP type, which is also suggested in ref. 32.

To investigate the spin relaxation in BLG quantitatively, we perform a detailed measurement of the gate voltage dependence of a BLG spin valve (device C) at 4 K. In figure 4a, $\tau_s$ and $D$ exhibit opposite dependences as a function of gate voltage, indicating the importance of DP spin relaxation. Quantitatively, it is known that the scattering rate for EY spin relaxation scales as $\frac{1}{\tau_s^{EY}} \sim \frac{1}{\tau_p} \sim \frac{1}{D}$, while the scaling for DP spin relaxation is $\frac{1}{\tau_s^{DP}} \sim \tau_p \sim D$ [7, 15, 25-28, 31]. Hence, if both EY and DP spin scattering are present, the spin lifetime is:

$$\frac{1}{\tau_s} = \frac{1}{\tau_s^{EY}} + \frac{1}{\tau_s^{DP}} = \frac{K_{EY}}{D} + K_{DP}D \qquad (2)$$

Figure 4b shows the spin relaxation rate ($1/\tau_s$) as a function of $D$ for BLG (device C at 4 K). The best fit by equation 2 yields $K_{EY} = 0.05 \pm 0.01$ ($10^{-2}$ m$^2$s$^{-1}$) ns$^{-1}$, and $K_{DP} = 1.24 \pm 0.09$ ($10^{-2}$ m$^2$s$^{-1}$)$^{-1}$ ns$^{-1}$ The contributions from EY spin relaxation are shown by the blue dashed line, and the contributions from DP spin relaxation are shown by the red dashed line. For the experimental range of $D$, the DP contribution to spin relaxation is much stronger than the EY contribution. For comparison, we plot the spin relaxation rate ($1/\tau_s$) as a function of $D$ (figure 4b inset) for SLG (device A at 4 K). The best fit parameters are $K_{EY} = 3.05 \pm 0.35$ ($10^{-2}$ m$^2$s$^{-1}$)ns$^{-1}$, and $K_{DP} = -0.02 \pm 0.10$ ($10^{-2}$ m$^2$s$^{-1}$)$^{-1}$ns$^{-1}$, which is zero within the error bars.

It is noted that longer spin lifetimes are observed in BLG (up to 6.2 ns) than in SLG (up to 1.0 ns). Theoretically, the intrinsic spin-orbit coupling in BLG is an order of magnitude larger than in SLG, which predicts shorter spin lifetimes for BLG [30]. The opposite experimental trend verifies that the spin relaxation in graphene is of extrinsic origin and the SLG is more sensitive to the extrinsic spin scattering than BLG. Possible sources of extrinsic EY spin relaxation include long-range (Coulomb) impurity scattering and short-range impurity scattering [13], while an extrinsic DP spin relaxation could arise from curvature of the graphene film [1, 5]. The transition from EY-dominated SLG to the DP-dominated BLG could be due to a strong reduction of the EY contribution because of enhanced screening of the impurity potential in thicker graphene [15, 33] and the smaller surface-to-volume ratio. However, a quantitative explanation for the substantial differences in spin relaxation between SLG and BLG will require further theoretical and experimental studies. Specifically, understanding the relationship between spin relaxation and the characteristics that differentiate SLG from BLG (e.g. band structure, lattice symmetry, bandgap formation in BLG, etc.) may be essential. For example, it has been shown by Dugaev *et al.* that in the presence of random spin-orbit interactions (which could be produced by curvature domains), the DP-like contribution to the spin lifetime has different values and decreases more rapidly with carrier concentration for massive fermions (BLG) than for massless fermions (SLG) [34, 35], which could contribute to the large changes in the fitting parameters $K_{EY}$ and $K_{DP}$.

In summary, spin relaxation in SLG and BLG spin valves has been investigated. By studying the spin lifetime and diffusion coefficient in SLG and BLG as a function of temperature and carrier concentration, contrasting behaviors are observed. For SLG, the EY spin scattering (e.g. impurity scattering) is dominant at low temperatures leading to the linear scaling of $\tau_s$ and $\tau_p$. For BLG, the temperature dependence shows an opposite trend between the spin lifetime and momentum scattering time, and the low temperature gate voltage dependence shows an inverse relationship of the two quantities, which indicate the dominance of DP spin relaxation.

We acknowledge technical assistance and discussion with E. Y. Sherman, K. M. McCreary, H. Wen, J. J. I. Wong, A. G. Swartz, K. Pi and Y. Li and the support of ONR (N00014-09-1-0117), NSF (DMR-1007057), and NSF (MRSEC DMR-0820414).

**Figure Captions:**

Fig. 1. (a) Nonlocal MR measurement of device A (SLG) at 300 K. The red (black) curve is measured as the field is swept up (down). E1 and E4 are Au electrodes (labeled 1 and 2), and E2 and E3 are Co electrodes (labeled 2 and 3). The large change in $R_{NL}$ indicated by the red arrow is due to the injection and transport of spin from E2 to E3. (b) Hanle measurements of device A at 300 K. The red (black) circles are data taken for parallel (antiparallel) Co magnetizations. The solid lines are the best fit by equation 1. (c) Spin lifetime (squares) and diffusion coefficient (circles) as a function of gate voltage at 300 K. (d) Spin lifetime (squares) and diffusion coefficient (circles) as a function of gate voltage at 4 K.

Fig. 2. Temperature dependence of SLG spin valves (device A). (a) Temperature dependence of spin lifetime at different gate voltages relative to the charge neutrality point. (b) Temperature dependence of diffusion coefficient at different gate voltages relative to the charge neutrality point. (c-f) Plot of spin lifetime vs. diffusion coefficient at T = 4 K, T = 10 K, T = 50 K and 100 K, T = 200 K and 300 K, respectively. The dotted line is a linear fit of the spin lifetime vs. diffusion coefficient.

Fig. 3. Gate and temperature dependence of BLG spin valves (device B). (a) Hanle measurement at 20 K for $V_g = V_{CNP}$. Inset: Hanle measurement at 300 K for $V_g = V_{CNP}$. (b) Spin lifetime (squares) and diffusion coefficient (circles) as a function of gate voltage at 300 K. (c) Spin lifetime (squares) and diffusion coefficient (circles) as a function of gate voltage at 20 K. (d) Temperature dependence of spin lifetime at different gate voltages relative to the charge neutrality point (e) Temperature dependence of the diffusion coefficient at different gate voltages relative to the charge neutrality point.

Fig. 4. Gate voltage dependence of BLG spin valves (device C). (a) Spin lifetime (squares) and diffusion coefficient (circles) as a function of gate voltage at 4 K. (b) Spin relaxation rate as a function of diffusion coefficient. The black solid line is the best fit based on equation 2. The dashed red (blue) line is the contribution of DP (EY) spin relaxation. Inset: Spin relaxation rate as a function of diffusion coefficient for SLG (device A) at 4K with the best fit (solid line) based on equation 2.

Figure 1

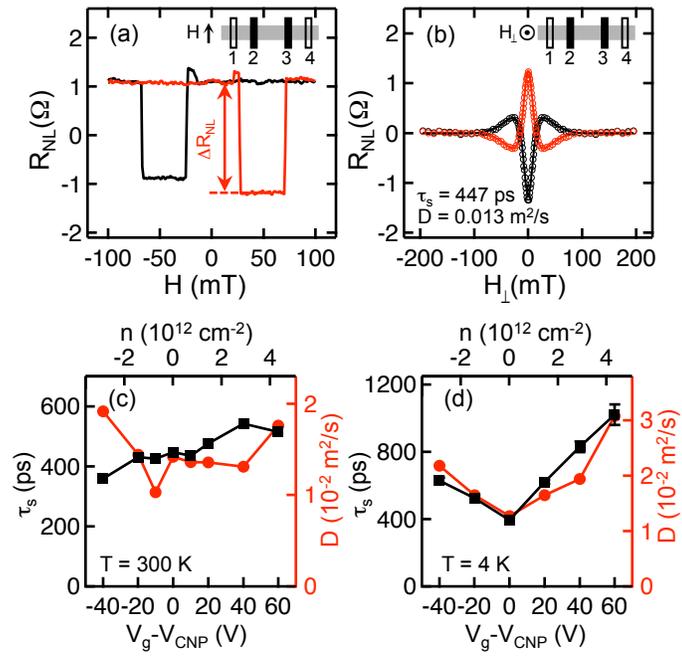

Figure 2

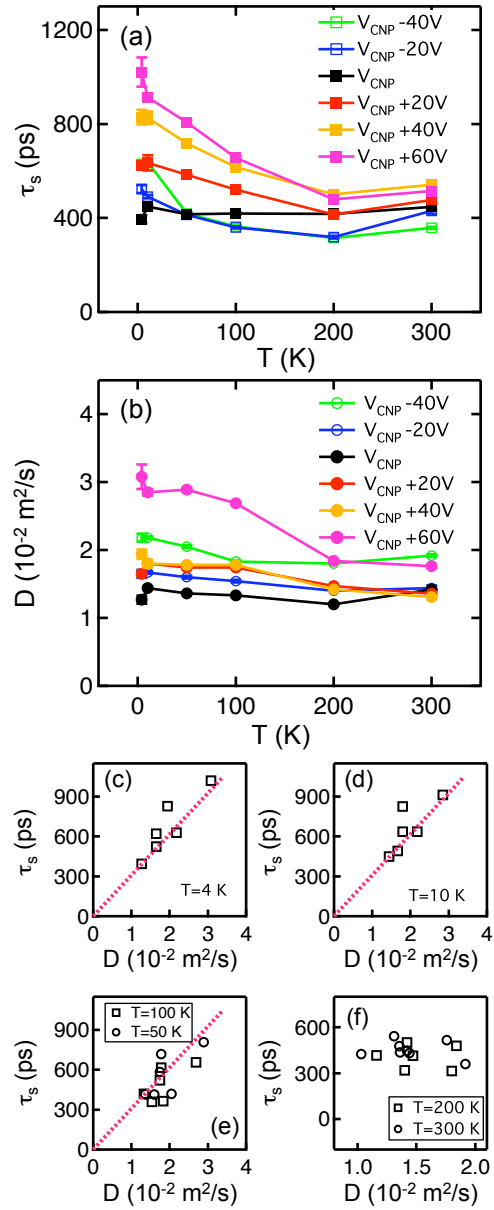

Figure 3

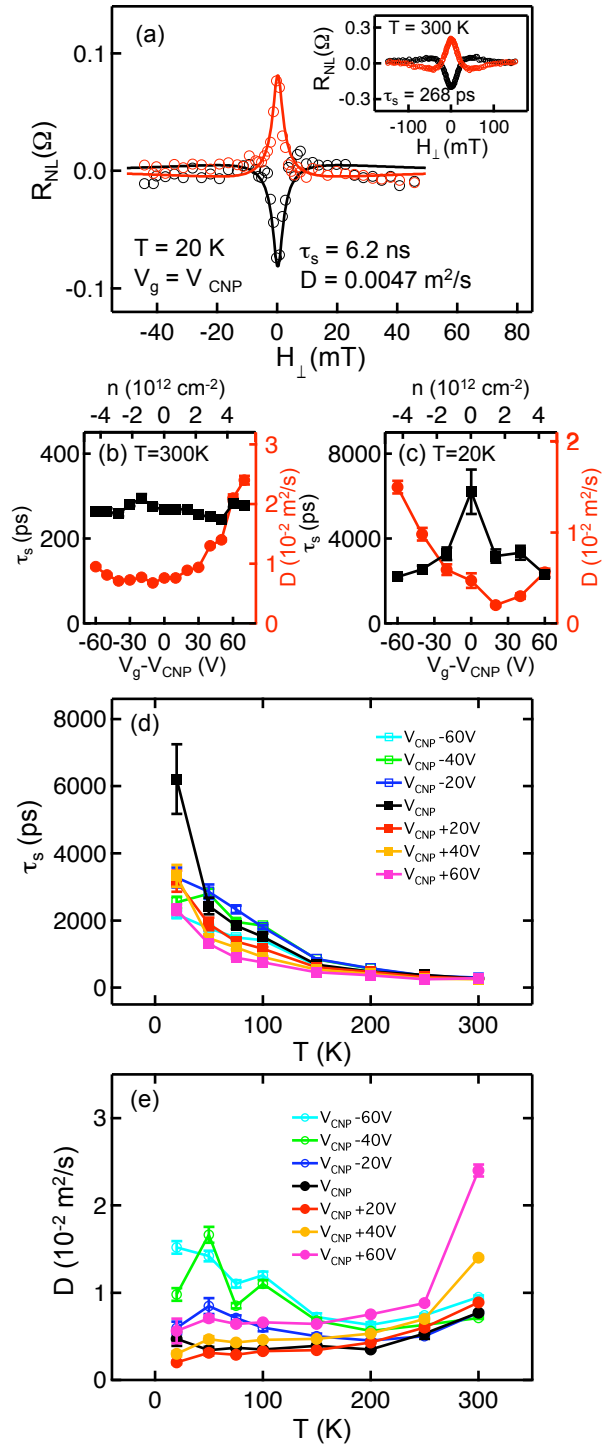

Figure 4

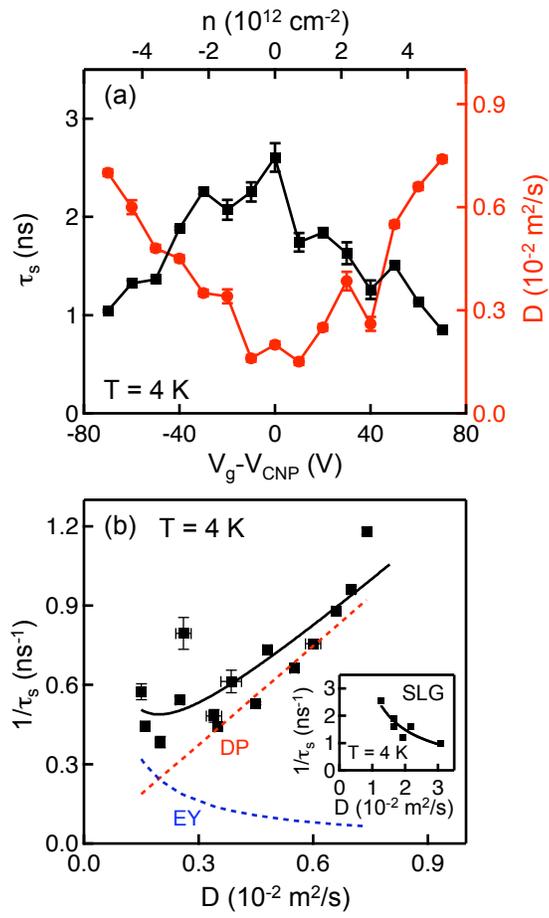

# Spin Relaxation in Single Layer and Bilayer Graphene

Wei Han and R. K. Kawakami[†]

Department of Physics and Astronomy, University of California, Riverside, CA 92521

[†] e-mail: roland.kawakami@ucr.edu

**Online Supplementary Information Content:**

1. The fabrication of graphene spin valves.

2. Nonlocal spin transport measurement.

`

## 1. The fabrication of graphene spin valves.

The graphene spin valves are fabricated using standard e-beam lithography and molecular beam epitaxy growth. The detailed steps are in the following:

**1) Graphene identification:** The graphene flakes are mechanically exfoliated from HOPG (SPI, GRADE ZYA) onto $SiO_2$ (300 nm thickness)/Si substrate (circled in black dashed line in figure S1a and S1b). The size of the graphene is typically 2-4 μm wide by 10-20 μm long. The SLG and BLG are identified by optical microscopy and Raman spectroscopy.

**2) Gold contact:** First, alignment marks are written into the PMMA/MMA bilayer resist using standard e-beam lithography (Leo SUPRA 55 with electron-beam pattern generation capability) (figure S1b). After development in MIBK:IPA solutions, optical pictures are taken and the device is designed using the software Design CAD. Using the align marks via the color contrast of $SiO_2$ and PMMA resist, the Au contacts are defined using a second step e-beam lithography with the same PMMA/MMA resist. Then, the devices are developed in MIBK:IPA solution and 5 nm Ti (used as the adhesion layer) and 60 nm Au are evaporated in an e-beam evaporator (Temescal BJD) with a base pressure of ~ $5\times10^{-6}$ torr. Finally, the lift-off process is done using PG remover at a temperature of 70 °C on a hot plate. At this stage, the device looks like figure S1c and S1d.

**3) Cleaning:** The device is annealed on a heater at 150 °C (heater temperature) for 1 hour in vacuum (~ $3\times10^{-9}$ torr) to clean the graphene surface.

**4) Cobalt tunneling contact:** Several Co contacts (between the two Au contacts) are also defined using standard e-beam lithography with PMMA/MMA bilayer resist. A small undercut will be generated in the bilayer resist for angle evaporation (figure S2a). After e-beam lithography, the sample is loaded into the MBE system with water-cooled manipulator. First,

`

0.12 nm of Ti is deposited at an angle of 9° and is converted to $TiO_2$ by exposure to oxygen partial pressure of $5\times10^{-8}$ torr for 30 min. MgO is deposited using an e-beam evaporator (MDC Corporation) and a single crystal MgO (MTI Corporation) as the source material. First, 0.8 nm layer of MgO is deposited at 9° for the tunnel barrier and then 3 nm of MgO is deposited at 0° for a masking layer. 80 nm of Co is deposited at 7° for the ferromagnetic electrodes. After all of these growths in the same MBE chamber, the devices are moved out and quickly loaded (<10 min) into an e-beam evaporator (Temescal BJD), and capped with 5 nm of $Al_2O_3$ on the Co to avoid further oxidation. . Finally, the lift-off process is done using PG remover at a temperature of 70 °C on a hot plate. The widths of the Co electrodes are varied between 80 nm and 300 nm to have different coercivities. At this stage, the device looks like figure S1e and figure S1f.

**2. Nonlocal spin transport measurement.**

Graphene spin valves are measured in a variable temperature $He^4$ cryostat with superconducting magnet.

**1) Wire bonding.** The device (big contact pad: 150 μm×150 μm) is first wire bonded (West Bond) to a homemade sample holder. The bonding wire is able to conduct to the Co pads through the $Al_2O_3$ cap layer.

**2) IV test.** After loading the device into the cryostat, we first measure the IV curve of the contacts. The geometry of the tunneling contact is shown in figure S2b. The differential contact resistance is highly non-linear (as shown in figure S2c).

**3) Nonlocal magnetoresistance (MR) measurement.** The geometry of the non-local spin transport measurement is shown in Figure S3a. Electrodes 1 and 4 are Au electrodes, while electrodes 2 and 3 are Co electrodes with different widths. Nonlocal voltages ($V_{NL}$) are measured using lock-in detection with an ac injection current of I = 1 μA rms at 13 Hz. The current is generated by a homemade voltage-controlled current source based on the op-amp Howland current source. Since the nonlocal voltage is usually small, it is pre-amplified 100 times using a Stanford SR560 voltage pre-amplifier. The relevant spin-injection occurs at electrode 2, where the carriers are driven from the Co into the SLG to generate a spin density within the SLG. This spin density then propagates toward electrode 3 via spin diffusion without a net charge current. The resulting spin-polarization beneath electrode 3 is then detected by a voltage measurement across electrodes 3 and 4 due to the spin-sensitivity of the Co electrodes. To extract the spin-signal from the background level, an external magnetic field oriented along the long axes of the Co electrodes is swept through the hysteresis loops of the Co electrodes to achieve parallel and anti-parallel magnetization alignments of the Co electrodes (figure S3b). Figures S3d shows representative scans of the non-local voltage as a function of applied magnetic field. As the field is swept up from negative field (red curve), the magnetizations of the Co electrodes are both negative and their parallel orientation leads to the higher value of the nonlocal voltage. At ~25 mT, the first Co electrode switches to achieve an antiparallel state, leading to a lower value of the nonlocal voltage. At ~70 mT, the other Co electrode switches to return the device to the parallel state. For the down sweep (black curve), similar behavior is observed with magnetization reversals occurring at negative field. The nonlocal resistance is the nonlocal voltage divided by the AC current. Figure S3e shows a nonlocal magnetoresistance curve with a constant background subtracted out. The nonlocal MR (or $\Delta R_{NL}$) is indicated by the arrow and its presence indicates that spins are injected at electrode 2 and transported via spin diffusion to electrode 3.

**3) Hanle spin precession measurement.** The Hanle effect is a very useful measurement, which provides an independent measure of the spin diffusion length and also yields the values of the spin lifetime and diffusion constant. To perform this, the parallel or anti-parallel state has to be achieved first. Then, the in-plane magnetic field is swept to 0, and the sample is rotated 90 degrees to align the magnetic field perpendicular to the device (figure S3c). This out-of-plane magnetic field ($H_\perp$) that induces spin precession at a Larmor frequency of $\omega_L = g\mu_B H_\perp/\hbar$, where $g$ is the g-factor, $\mu_B$ is the Bohr magneton, and $\hbar$ is reduced Planck's constant.

Figure S1

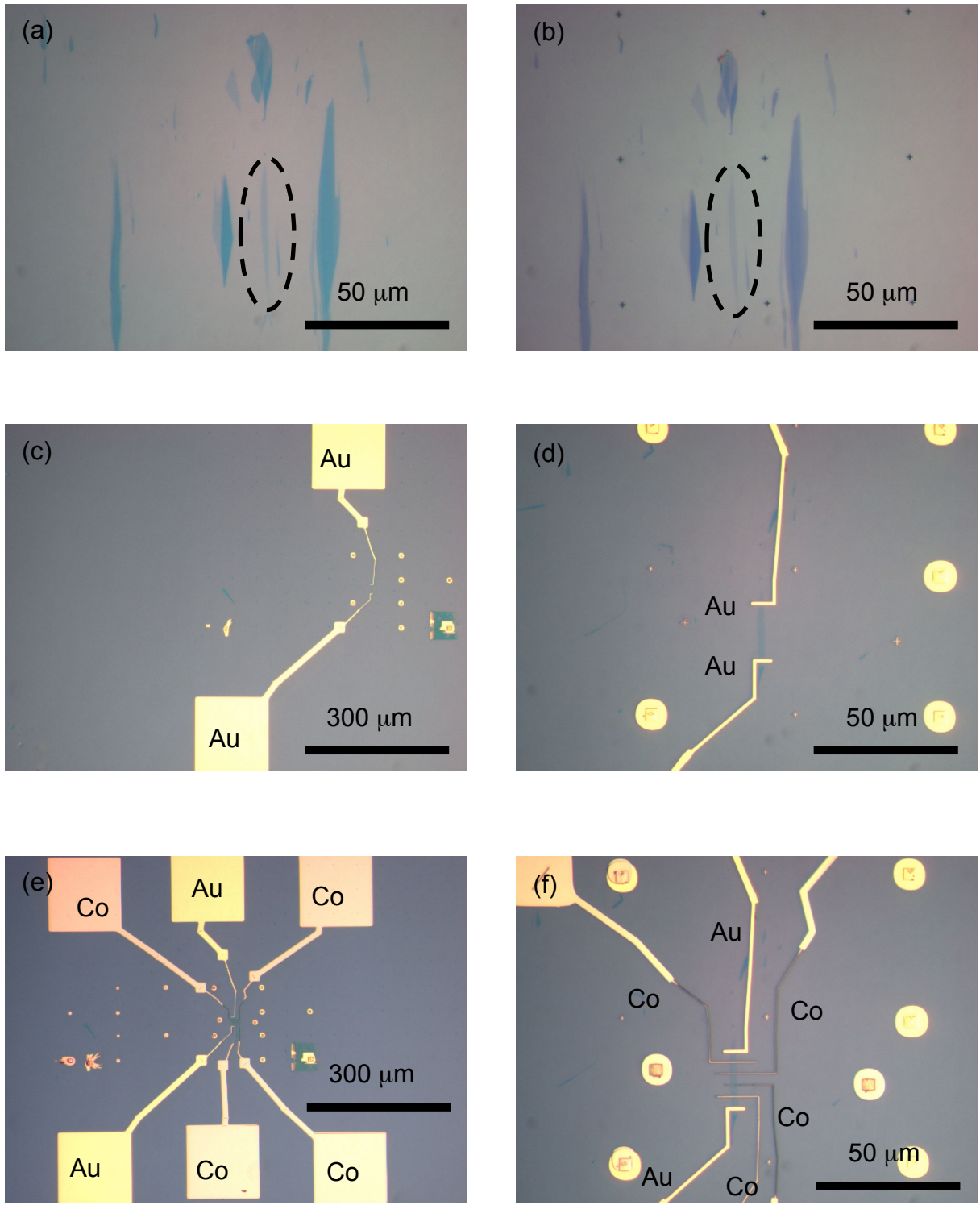

Figure S1, Device fabrication.
(a) Graphene on SiO$_2$/Si; (b) Graphene on SiO2/Si with alignment marks on PMMA/MMA bilayer resist; (c-d) Graphene spin valves with Au contacts. (e-f) Graphene spin valves with Co electrodes.

Figure S2

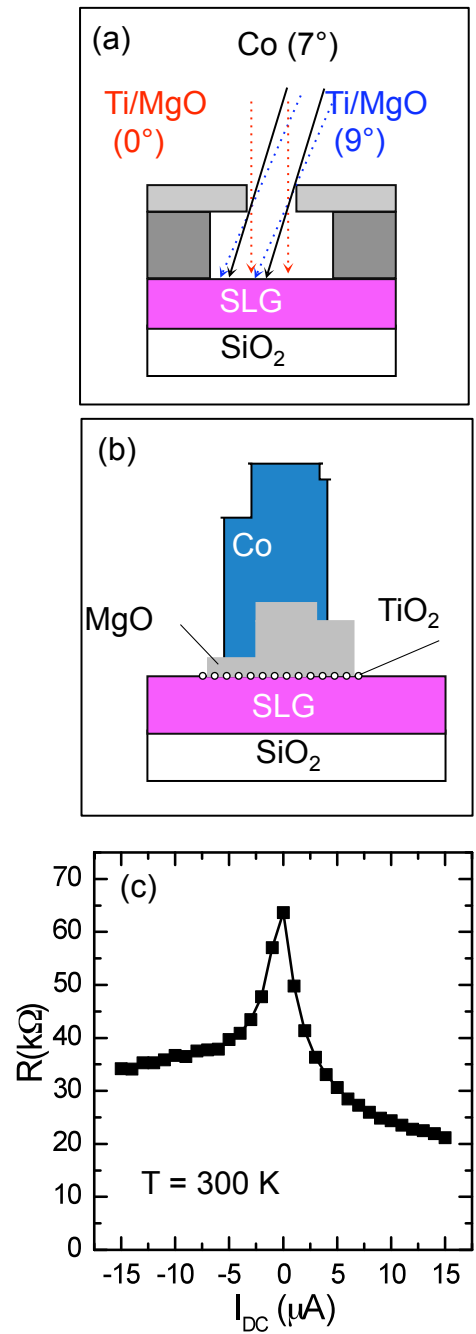

Figure S2, tunneling contacts Co/MgO (TiO$_2$)/graphene.
(a) Angle evaporation of Ti, MgO and Co; (b) Tunneling contacts Co/MgO (TiO$_2$)/graphene; (c) Typical differential contact resistance of the tunneling contact at 300 K.

Figure S3

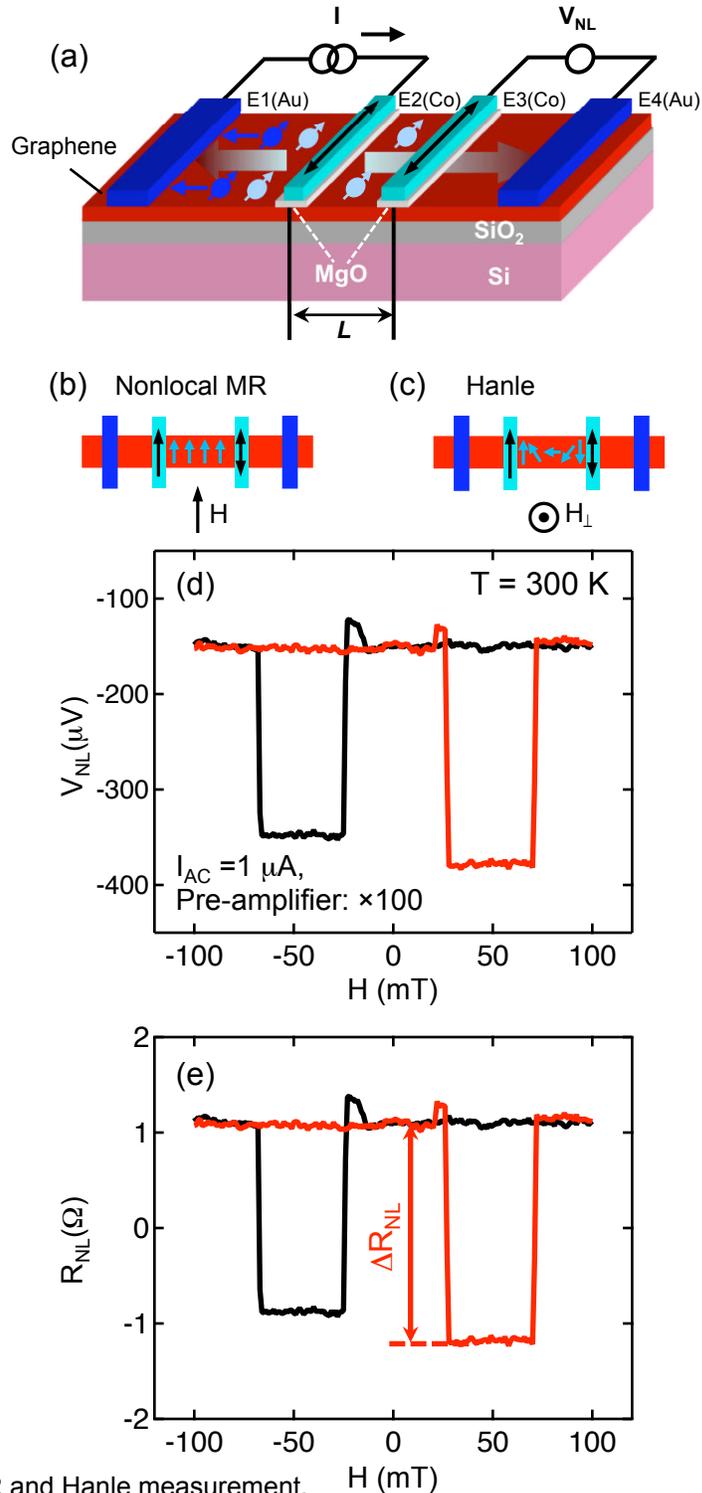

Figure S3, Nonlocal MR and Hanle measurement.
(a) Device structure and nonlocal MR measurement of the spin diffusion; (b) Nonlocal MR measurement; (c) Hanle measurement; (d) Raw data of nonlocal voltage as the magnetic field is swept up and down; (d) the corresponding nonlocal MR curve. The arrow indicates the nonlocal MR.